\documentstyle[fleqn,epsf,11pt]{article}
\def\bo{{\raise.15ex\hbox{\large$\Box$}}}

\def\dag{^{\dagger}{}}
\def\rarr{\rightarrow}

\def\ordless{{\lower2mm\hbox{$\,\stackrel{\textstyle <}{\sim}\, $}}}
\def\ordmore{{\lower2mm\hbox{$\,\stackrel{\textstyle >}{\sim}\, $}}}

\newtoks\slashfraction
\slashfraction={.13}
\def\slash#1{\setbox0\hbox{$\, #1$}
\setbox0\hbox to \the\slashfraction\wd0{\hss \box0}/\box0}

\def\leftrightarrowfill{$\mathsurround=0pt \mathord\leftarrow \mkern-6mu
        \cleaders\hbox{$\mkern-2mu \mathord- \mkern-2mu$}\hfill
        \mkern-6mu \mathord\rightarrow$}
\def\overleftrightarrow#1{\vbox{\ialign{##\crcr
        \leftrightarrowfill\crcr\noalign{\kern-1pt\nointerlineskip}
        $\hfil\displaystyle{#1}\hfil$\crcr}}}
\def\startarray{\left( \begin{array}}
\def\finarray{\end{array} \right)}
\def\starteq{%\vspace{.1in}
\begin{eqnarray}}
\def\fineq{\end{eqnarray}
}

\catcode`@=11
\def\underline#1{\relax\ifmmode\@@underline#1\else
$\@@underline{\hbox{#1}}$\relax\fi}
\catcode`@=12
 
\newskip\humongous \humongous=0pt plus 1000pt minus 1000pt

\newif\ifdtup
 
\def\textcite#1{Ref.~{\cite{#1}}}

\def\thefootnote{\fnsymbol{footnote}}
 
\def\author#1#2{{\bf #1} \\ {\em #2}\vspace{5mm}}

\def\bold#1{\setbox0=\hbox{$#1$}%
     \kern-.025em\copy0\kern-\wd0
     \kern.05em\copy0\kern-\wd0
     \kern-.025em\raise.0433em\box0 }

\font\myninerm=cmr9 
\font\myninebf=cmbx9

\newdimen\bigdimentete
\newdimen\dimentete
\dimentete = \bigdimentete
\def\mynegskip{\vskip -4pt plus 1pt minus 1 pt}

\def\ftoday{ {\number\day \space\ifcase\month 
\or Janvier\or F\'evrier\or Mars\or Avril\or Mai
\or Juin\or Juillet\or Ao\^ut\or Septembre\or Octobre
\or Novembre \or D\'ecembre\fi \number\year}}    
\def\today{ {\ifcase\month 
\or January\or February\or March\or April\or May
\or June\or July\or August\or September\or October
\or November \or December\fi \space\number\day, \number\year}}
\newbox\entetebox
\setbox\entetebox=\vbox{
\hbox to \dimentete{\myninerm\hfil ECOLE POLYTECHNIQUE\hfil}%should be ninerm?
\hbox to \dimentete{\myninebf\hfil CENTRE DE PHYSIQUE THEORIQUE\hfil}
\hbox to \dimentete{\myninerm\hfil 91128 PALAISEAU CEDEX-FRANCE\hfil}
\mynegskip
\hbox to \dimentete{\hfil\hbox{\vrule width10em height0.6pt}\hfil}
\mynegskip
\hbox to \dimentete{\myninerm\hfil T\'el.~: (33) (1) 
69 33 47 36 \hfil}%should be sevenrm?
\mynegskip
\hbox to \dimentete{\myninerm\hfil T\'elex~: ECOLEX 601 596 F\hfil}
\mynegskip
\hbox to \dimentete{\myninerm\hfil Fax.~: (33) (1) 69 33 30 08\hfil}
\mynegskip
\hbox to \dimentete{\myninerm\hfil  CNRS UMR 7644\hfil}
}
\def\sendadr{\null\vskip0.5cm
\noindent
\hskip -8em
\hbox to \bigdimentete{\copy\entetebox}} 

\def\title#1#2#3#4#5{\thispagestyle{empty}
        \begin{center} \vspace*{1cm} { \bf #3} \\[.5in] {#4{}}
        \end{center} \vfill \centerline{ ABSTRACT}
   {\nopagebreak \noindent\begin{quotation}\noindent {\small #5}
   \end{quotation}} \vfill {#2} \hfill\begin{tabular}{r} {#1} 
        \end{tabular}  \newpage
        \def\thefootnote{\arabic{footnote}}}
\def\prefer{\section*{}
    \list{[\arabic{enumi}]}{\usecounter{enumi}\settowidth\labelwidth{[000]}
      \leftmargin\labelwidth\advance\leftmargin\labelsep \rightmargin=0pt}
        \small \sfcode`\.=1000\relax}

\def\refer#1{\section*{\large \sc {#1}}
    \list{\arabic{enumi}.}{\usecounter{enumi}\settowidth\labelwidth{[000]}
      \leftmargin\labelwidth\advance\leftmargin\labelsep \rightmargin=0pt}
        \raggedright \small \sfcode`\.=1000\relax}

\def\ReFer#1#2{\section*{\large\sc#1}
    \list{[\arabic{enumi}]}{\usecounter{enumi}\settowidth\labelwidth{#2}
      \leftmargin\labelwidth\advance\leftmargin\labelsep \rightmargin=0pt}
        \raggedright \small \sfcode`\.=1000\relax}

\def\REFER#1#2{\section*{\large\sc#1}
    \list{#2 {enumi}.}{\usecounter{enumi}\settowidth\labelwidth{[000]}
      \leftmargin\labelwidth\advance\leftmargin\labelsep \rightmargin=0pt}
        \raggedright \small \sfcode`\.=1000\relax}

\def\startbib{\vspace{1in}\begin{refer}{References}
\small\frenchspacing\nopagebreak}
\def\endbib{\end{refer} \normalsize \nonfrenchspacing}
\def\startfig{\newpage \centerline{{\sl Figure captions}} \begin{itemize}}

\def\endfig{\end{itemize}}
 
\newcommand{\be}{\begin{equation}}
\newcommand{\ee}{\end{equation}}
\newcommand{\bea}{\begin{eqnarray}}
\newcommand{\eea}{\end{eqnarray}}

\newcommand{\AmS}{{\protect\the\textfont2
  A\kern-.1667em\lower.5ex\hbox{M}\kern-.125emS}}
\begin{document}
% declarations for front matter
\title{January 2003} {PC073.0103} { CP-violating asymmetry in $B^- \rarr \pi^+ \pi^- K^-$
   and  $B^- \rarr K^+ K^- K^-$    decays  
\footnotetext{${\dag}$  
Talk given at the Quark Confinement and Hadron Spectrum V , Gargnano,Italy 
10-14 September 2002}}
{\author{T. N. PHAM}
{Centre de Physique Th\'eorique, \\
Centre National de la Recherche Scientifique, UMR 7644, \\  
Ecole Polytechnique, 91128 Palaiseau Cedex, France}}        
{The recent Belle and Babar measurements of the
branching ratios $B^- \rarr \pi^+ \pi^- K^-$
and $B^- \rarr K^+ K^- K^-$ and $B^\pm \rarr \chi_{c0} K^\pm$ have renewed
interests in these decays as another mean to look for direct
CP violation in $B$ decays. In this talk, I would like to discuss 
a recent analysis of the CP violating  asymmetry in the partial widths
for the decays  $B^- \rarr \pi^+ \pi^- K^-$
and $B^- \rarr K^+ K^- K^-$, which results from the interference of the
non resonant amplitude with the resonant amplitude
$B^\pm \rarr \chi_{c0} K^\pm $ $ \rarr\pi^+ \pi^- K^\pm  $ and
$B^\pm \rarr \chi_{c0} K^\pm$ $ \rarr K^+ K^- K^\pm$.
For  $\gamma \simeq 58^o$ we predict that the 
partial width asymmetry could reach  $10 \%$ for 
the $B^- \rarr \pi^+ \pi^- K^-$ decay
and $16\%$ for the  $B^- \rarr K^+ K^- K^-$decay.}

In $B$ decays into 3 light mesons, for example, in 
$B^- \rarr \pi^+ \pi^- \pi^-$ decays, CP asymmetry could 
arise from the interference of the non resonant amplitude 
with the resonant amplitude coming from the decays 
$B^{-}\rarr \chi_{c0}\pi^{-}$ $\rarr$ $\chi_{c0}\rarr \pi^- \pi^-$ 
\cite{Eilam,Deshpande} and an  
estimate for the partial width asymmetry at the $\chi_{c0}$
resonance  was given recently\cite{Fajfer}. 
As there is no theoretical prediction
 for the $B^{-}\rarr \chi_{c0}\pi^{-}$
decay rate, no firm prediction for the asymmetry could be given. It
is now possible to make a similar
analysis for the CP asymmetries in the $B^-\rarr \pi^+ \pi^- K^- $ 
and $B^-\rarr K^+ K^- K^- $ decays using the recent 
Belle Collaboration measurements\cite{Belle1,Belle2}~:
${\rm BR}(B^-\rarr \chi_{c0} K^-) = (6.0^{+2.1}_{-1.8})\times 10^{-4} $,
${\rm BR}(B^-\rarr K^+ K^- K^-) = (37.0\pm 3.9 \pm 4.4) \times 10^{-6} $
and 
${\rm BR}(B^-\rarr \pi^+ \pi^- K^-) =(58.5\pm 7.1 \pm 8.8) \times 10^{-6} $,
and the  value 
${\rm BR}(B^-\rarr \chi_{c0} K^-) = (2.4\pm 0.7)\times 10^{-4} $
from the Babar Collaboration\cite{Babar}. 
In this talk, I would like
to discuss a recent work\cite{Fajfer1} on the CP asymmetries 
to look for direct CP violation in these decays. 
I will present only the main point as details are given in the
published work.

The effective weak Hamiltonian  for the nonleptonic Cabibbo-suppressed
$B$ meson decays is given by\cite{Ali,Deshpande1,Isola}
%\newpage
\begin{eqnarray}
{\mathcal H}_{\rm eff}&  = & \frac{G_F}{{\sqrt 2}} [V_{us}^* V_{ub}
(c_1 O_{1u} +
c_2 O_{2u} )  + V_{cs}^* V_{cb} (c_1 O_{1c} +c_2 O_{2c} ) ]\nonumber\\
& - &\sum_{i=3}^{10} ([V_{ub} V_{ud}^* c_i^u
     +  V_{cb} V_{cs}^* c_i^c + V_{tb} V_{ts}^* c_i^t) O_i ] + h.c.
\label{eq1}
\end{eqnarray}
where $O_{1}, O_{2}$ are the usual tree-level operators and $O_{3}-O_{6}$
are the penguin operators. As usual with the factorization model we use
in our analysis, the hadronic marix elements are obtained from the 
effective Hamiltonian in 
Eq.(\ref{eq1}) with $c_{i}$ replaced by $a_{i}$ where $c_{i}$
are next-to-leading Wilson coefficients. Since  $a_3$ and $a_5$
are one order of magnitude smaller than $a_4$ and $a_6$, the contributions
from $O_{3}$ and $O_{5}$ can be safely neglected. For
$N_{c}=3$, $m_b = 5 \rm \, GeV$, we use\cite{Deshpande1,Isola}~: 
$a_1= 1.05$ ,  $a_2=  0.07$ , $ a_4=-0.043- 0.016\,i$, 
$a_6=  -0.054- 0.016\,i$.
The matrix element of $O_{1}$ for the 
non resonant $B ^- \rarr \pi^+ \pi^- K^-$ amplitude is then given by
\begin{eqnarray}
&&<O_{1}>_{\rm nr} = < K^-(p_3) \pi^+(p_1) \pi^-(p_2)| O_1| B(p_B)>_{nr} = 
-[f_3 m_3^2 r^{nr} \nonumber\\
&&+ \frac{1}{2} f_3(m_B^2 - m_3^2 -s)w_+^{nr} + 
\frac{1}{2}f_3 (s + 2 t - m_B^2 - 2 m_1^2 - m_3^2)w_-^{nr}]
\label{o1mv}
\end{eqnarray}
where  $s= (p_B - p_3)^2$, $t= (p_B - p_1)^2$ and $u= (p_B-p_2)^2$ and 
the form factors $w_\pm^{nr}$  and $r^{nr}$ in 
$< \pi^-(p_1) \pi^+(p_2) | ({\bar u} b)_{V -A}| B^- (p_B)>_{nr}$ 
are computed in the Heavy Quark Effective Theory
(HQET)\cite{Fajfer,Bajc2} with the $B, B^{*}$ propagators in the
pole contributions are the full propagators.
The operator $O_4$ has the same kind of decomposition as $O_1$,
while  $O_6$ can be rewritten as the product of 
density operators. We have, in the factorization model: 
\begin{eqnarray}
&&<O_{6}>_{\rm nr} =< K^-(p_3) \pi^+(p_1) \pi^-(p_2)| O_6| B(p_B)>_{nr} 
= -(\frac{ {\mathcal B}}{m_B})[ 2 \frac{f_1 f_2}{f_3} m_3^2 r^{nr}\nonumber\\
&&+ \frac{f_1 f_2}{f_3} (m_B^2 + m_3^2 -s)w_+^{nr} 
+ \frac{f_1 f_2}{f_3} (s + 2 t - m_B^2 - 2 m_1^2- m_3^2 )w_-^{nr}].
\label{o6mv}
\end{eqnarray}
The $B^- \rarr K^- \pi^+ \pi^-$ decay amplitude is 
\be
\kern -0.5cm    {\mathcal M}_{nr} = \frac{G}{{\sqrt 2}}[ V_{ub} V_{us}^* a_1
<O_{1}>_{\rm nr} - \ V_{tb} V_{ts}^* ( a_4<O_{4}>_{\rm nr}
   + \ a_6<O_{6}>_{\rm nr} )].
\label{amp}
\ee
from which we obtain  the non resonant branching ratios:
\be
{\rm BR}(B^-\rarr K^- \pi^+ \pi^-)_{nr} = T + P + I_1 \cos \gamma
   + I_2 \sin \gamma. 
\label{kpp-b1}
\ee
with $\gamma$ the CP violating weak phase. 
$T = 7.0\times 10^{-6}$, $P= 7.5 \times 10^{-5}$,
$I_1 = -4.3 \times 10^{-5}$, $I_2 = -1.5 \times 10^{-5}$, and 
\be
{\rm BR}(B^-\rarr K^- K^+ K^-)_{nr} = T + P + I_1 \cos \gamma
   + I_2 \sin \gamma. 
\label{kkk-b1}
\ee
with $T = 3.4\times 10^{-6}$, $P= 3.7\times 10^{-5}$, 
$I_1 = -2.1 \times 10^{-5}$ ,$I_2 = -7.4 \times 10^{-6}$. 
These values should be considered as upper limit for the non resonant
branching ratios. The  CP asymmetry of the total decay rate is 
\be
A = \frac{ \sin \gamma N_1}{N_2 + \cos \gamma N_3}, 
\label{asint1}
\ee
with $N_1= -3.0 \times 10^{-5}$, $N_2= 16.4 \times 10^{-5}$ , 
$N_3= -8.6\times 10^{-5}$ for $B^- \rarr  \pi^+ \pi^- K^-$ decay
and $N_1= -1.5 \times 10^{-5}$, $N_2= 8.2 \times 10^{-5}$,
$N_3= -4.2\times 10^{-5}$ for $B^- \rarr  K^+ K^- K^-$ decay. This result is
essentially independent of the form factors since in the 
$SU(3)$ limit, $<O_{6}>= ({2\mathcal B}/m_{B})<O_{1}>$. 
The resonant contribution  is given by 
\begin{eqnarray}
&&{\mathcal M}_{r}(B^- \rarr  \chi_{c0} K^- \rarr \pi^+ \pi^- K^-)  = \nonumber\\
&&{\mathcal M}(B^{-} \rarr \chi_{c0} K^- ) \frac{1}{ s - m_{\chi_{c0}}^2 +  i
\Gamma_{\chi_{c0}} m_{\chi_{c0}}}
{\mathcal M}( \chi_{c0} \rarr \pi^+ \pi^- ).
\label{ares}
\end{eqnarray}
Using the Belle measured $B^- \rarr  \chi_{c0} K^- $ branching 
ratio\cite{Belle1}, we
computed the differential decay rates and CP asymmetries for
the two-pion and two-kaon system in the $\chi_{c0} $ mass region (see 
Figs.1 and Figs.2 of the published work\cite{Fajfer1}).
For the integrated CP asymmetry over the $\chi_{c0} $ resonance, we find
$ A_p(B^\pm\rarr K^\pm \pi^+ \pi^-) = 7.9 \sin \gamma/(73 - 1.2 \cos \gamma)$
and 
$A_p(B^\pm\rarr K^\pm K^+ K-) =  7.2 \sin \gamma/(41 - 5.6 \cos \gamma ) $ .

\bigskip

%\section*{Acknowledgments}

\end{document}